\documentclass[aps,pre,superscriptaddress,showpacs,twocolumn]{revtex4}
\usepackage{amsmath}
\usepackage{graphicx}
\begin{document}

\title{Effect of Quantum Fluctuations in an Ising System on Small-World
  Networks}

\author{Hangmo Yi}
\affiliation{Korea Institute for Advanced Study, 207-43
  Cheongryangri-dong Dongdaemun-gu, Seoul 130-012, Korea}
\author{Mahn-Soo Choi}
\affiliation{Korea Institute for Advanced Study, 207-43
  Cheongryangri-dong Dongdaemun-gu, Seoul 130-012, Korea}
\affiliation{Department of Physics, Korea University, 5-ka 1 Anam-dong
  Sungbuk-ku Seoul,136-701, Korea}

\begin{abstract}
We study quantum Ising spins placed on small-world networks.  
A simple model is considered in which the coupling between 
any given pair of spins is a nonzero constant if they are linked 
in the small-world network and zero otherwise.  
By applying a transverse magnetic field, we have investigated 
the effect of quantum fluctuations.  Our numerical analysis shows 
that the quantum fluctuations do not alter the universality 
class at the ferromagnetic phase transition, which is 
of the mean-field type.  The transition temperature is reduced by 
the quantum fluctuations and eventually vanishes at 
the critical transverse field $\Delta_c$.  With increasing 
rewiring probability, $\Delta_c$ is shown to be enhanced.  
\end{abstract}


\maketitle

Various phenomena in nature and human society can be understood in terms
of dynamical interaction among individual elements on a complex network.
The equilibrium and dynamic properties of such systems depend strongly
on the topology of the underlying network as well as how the individuals
interact with each other.  For example, the antiferromagnetic Ising
model suffers from frustration effects on a triangular lattice, whereas
it has a simple ground state on a square lattice.  For the mathematical
simplicity, the network is often regarded as either completely random or
purely regular.  In reality, however, most of the biological systems,
solid-state systems, and human societies lie somewhere between these
two extremes.  Furthermore, it has recently been shown that such
networks, now widely known as ``small-world'' networks, exhibit
mixed properties: some of which are common to completely random
networks, some common to purely regular networks, and others unique to
small-world networks \cite{Barabasi02a,Kochen89a}.

A simple yet inspiring mathematical model for a small-world network has
been proposed by Watts and Strogatz (WS) \cite{watts98a}.  One starts
from a regular network and ``rewires'' the links with probability
$p$.  As $p$ varies from $0$ to $1$, the resulting network
``interpolates'' from a purely regular network to a completely random
network.  The value of the model so generated comes from the fact that
it captures important physics in 
a wide range of physical systems, which can be
described by neither a regular nor a random network.

In recent years, many authors have studied 
scaling properties, crossover
behavior, percolation behavior \cite{newman99a,moore00b}, 
the spread of infectious diseases \cite{kuperman01a,moore00a},
signal-propagation speed \cite{watts98a}, computational power, 
and synchronization \cite{barahona02a,ito02a} 
of small-world networks.  More recently,
phase transitions of the Ising model \cite{barrat00a,gitterman00a,herrero02a} 
and $XY$ model \cite{kimBJ01a} on small-world
networks have also been studied, where the crossover from
one-dimensional to mean-field behavior has been found.  Although most of
the works mentioned above are concerned about classical statistical
problems on small-world networks, Zhu and
Xiong \cite{zhuCP00a,zhuCP01a} have recently studied the
localization-delocalization transition of electronic states on a
small-world network which allows a quantum mechanical hopping among
sites.

In this work, we adopt a transverse-field Ising model on small-world
networks, to investigate effects of the interplay between the unique
topology of small-world networks and the quantum fluctuations.  From
the previous works \cite{barrat00a,gitterman00a,herrero02a,kimBJ01a}, it
is known that the small-world network topology enhances the correlation
between the spins on the network and leads to the mean-field behavior of
the system.  Since the quantum fluctuations, introduced by the
transverse field in our model, tend to destroy correlations, one can
anticipate a non-trivial competition between the small-world network
topology and the quantum fluctuations.

Another motivation for studying the transverse-field Ising model on
small-world networks is provided by the recent wide interest in quantum
computing.  A quantum computer can be regarded as controllable quantum
spins (quantum mechanical two-state systems) interacting with each other
through a network \cite{nielsen00booka}.  In realistic circumstances, the
control of the spins, which can be achieved by carefully tuning 
the local magnetic field at each node and the coupling between each pair of
spins, is imperfect.  The effects of imperfections on quantum computing
has been studied on a completely random network and have been shown to 
give rise to computation errors, 
which grow fast (exponentially or polynomially) with the
number of quantum bits (or in short, qubits) \cite{songPH01a}.  
It is therefore
much worth studying the imperfection effects in terms of qubits on
small-world networks.  With the transverse field assumed uniform over
the whole network, the model in our work may not directly represent a
realistic quantum computer with imperfect controls, yet our work might
provide a stimulation for studies in that direction.

We consider $N$ interacting spins in a uniform transverse magnetic field 
$\Delta$.  We assume that the interaction between spins is Ising-type.
The Hamiltonian for the system is then given by
\begin{equation}
\label{model:H}
H = H_z + H_x =
- \sum_{i<j}J_{ij}\sigma_i^z\sigma_j^z
- \Delta\sum_{i=1}^N \sigma_i^x \,,
\end{equation}
where $\sigma^x$, $\sigma^y$, and $\sigma^z$ are the Pauli matrices.
The coupling $J_{ij}$ between the two spins at the $i$th and $j$th sites
are defined on a small-world network (see below for the precise way 
the network is constructed).  Namely, we regard that the spins are placed on 
the small-world network, and that $J_{ij}=J$ if $i$ and $j$ are linked on
the network and $J_{ij}=0$ otherwise \cite{uniform}.
The sign of $J$ is irrelevant in this model since 
the ferromagnetic ($J>0$) and the antiferromagnetic ($J<0$) 
models are interchangeable through a transformation 
that rotates every other spin about $x$-axis.  
Below, we will assume $J>0$.  

The small-world network is constructed in a very similar way as used by
WS: We take a one-dimensional regular network with $2k$ nearest
neighbors \cite{k>1} with a periodic boundary condition ($\sigma^l_{i+N}
\equiv \sigma^l_i$, $l=x,y,z$).  Among the total of $Nk$ links we
randomly choose $pNk$ of them.  One end of each chosen link is then
rewired to a random site keeping the other end as a pivot point.
Among the resulting networks, we discard those with disconnected parts,
i.e., we only considers a network where all the nodes belongs to a
single cluster.

Without going into details, several properties of this model may be
qualitatively understood as follows.  In the classical case
($\Delta=0$), this model is known to undergo a mean-field type
ferromagnetic phase transition at a finite temperature for arbitrarily
small $p>0$ \cite{barrat00a}.  Finite transverse magnetic field 
introduces quantum fluctuations, which compete with the correlations
enhanced by the topology of the small-world network.  As a consequence,
one expects $\Delta$ to suppress the transition temperature $T_c$.  In
the fully quantum mechanical case ($T=0$) on the nearest-neighbor
regular network ($k=1$), the Hamiltonian in Eq.~(\ref{model:H}) shows a
quantum phase transitions at $\Delta_c=1$ of the Kosterliz-Thouless
type \cite{sachdev98booka}.

Below we will calculate the phase boundary, i.e., the transition
temperature $T_c$ as a function of $\Delta$, using the quantum
Monte Carlo simulation.  
The universality class to which the phase transition belongs 
will also be determined from various critical exponents.  
Before we give the results, we summarize the quantum Monte Carlo
method specific to our model.

We divide the inverse temperature $\beta=k_BT$ into $M$ pieces with
spacing $\epsilon\equiv\beta/M$.  The
partition function $Z(\beta)=\mathrm{tr}\exp\left(-\beta{H}\right)$ is then
approximated by the Trotter production formula
\begin{equation}
\label{eq:Trotter1}
Z(\beta) \approx \mathrm{tr}\left[
  \exp\left(-\epsilon H_x\right)
  \exp\left(-\epsilon H_z\right)\right]^M \,,
\end{equation}
the error of which is on the order of $\epsilon^2$ ($M\to\infty$).
We now use the completeness relation
\begin{equation}
\sum_i\sum_{S_{i,t}=\pm1}|S_{i,t}\rangle\langle S_{i,t}|=1 \,,
\end{equation}
where $|S_{i,t}\rangle$ is the eigenstates of the $\sigma_i^z$, and put
it between the $t$th and $(t+1)$th temperature slices in
Eq.~(\ref{eq:Trotter1}).  We may then write
\begin{multline}
\label{eq:Z}
Z(\beta) = \sum_{\{S_{i,t}\}} \exp\left[
  \sum_{i<j}\sum_{t=1}^M \epsilon{J}_{ij}\,S_{i,t}S_{j,t}
\right. \\ \left. \mbox{}
  + A\sum_{i=1}^N\sum_{t=1}^M S_{i,t}S_{i,t+1}
  + NMB \right] \,,
\end{multline}
where
\begin{align}
A & = \frac{1}{2}\ln\coth(\epsilon\Delta) \,,\\
B & = \frac{1}{2}\ln\left[\cosh(\epsilon\Delta)\sinh(\epsilon\Delta)\right] \,.
\end{align}
Interpreting the temperature slices as imaginary time intervals, 
the above equation becomes a classical Ising model defined on a
$(1+1)$-dimensional lattice.  The coupling in the spatial direction
(indicated by the indexes $i$ and $j$) is given by $J_{ij}$ and
contains the topology of the small-world network.  The coupling in the
temporal direction (indicated by the index $t$), on the other
hand, is a nearest neighbor coupling.

Due to randomness in the topology of the network, this problem is 
naturally suited to a numerical analysis.  
We have thus performed a quantum Monte Carlo analysis using the above 
partition function.  In order to increase efficiency 
near and below the ferromagnetic transition temperature, 
a cluster algorithm was adopted \cite{newman01booka}.  
Since our model is different from a usual regular 
Ising model, we have developed a new cluster algorithm 
which takes into account the small-world network topology 
and the anisotropy between space and time axes.  

\begin{figure}
\centering
\includegraphics[width=8cm]{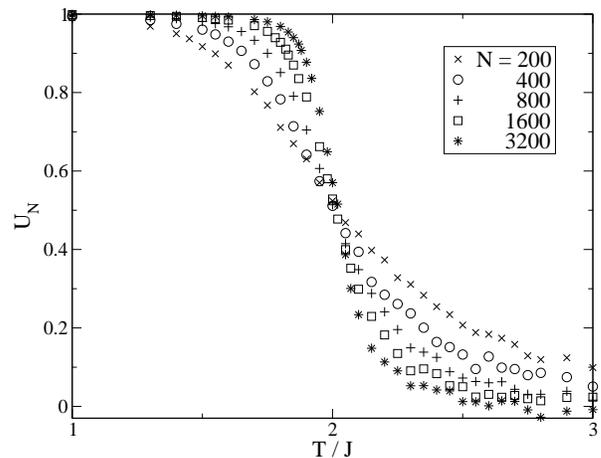}
\caption{The fourth order Binder cumulant $U_N$ is drawn 
as a function of temperature for five different 
system sizes $N$.  The parameters are given 
by $k=2$, $p=0.1$, and $B/J=1$.  
In the time direction, the size was 
fixed at $M=30$.  As $N$ grows, the crossing point 
converges to one single point, which is $T/J=2.04$ in 
this graph.}
\label{fig:binder}
\end{figure}

The transition temperature $T_c(\Delta)$ has been determined 
from a finite size scaling method.  Varying temperature, we 
computed the fourth order Binder cumulant \cite{binder88booka}
\begin{equation}
U_N(T) = \frac{1}{2} \left( 3-\frac{[\langle m^4\rangle]}{[\langle m^2\rangle^2]} \right)
\end{equation}
for several different values of $N$.  This quantity 
varies from one to zero while temperature is swept 
from zero (maximum order) to infinity (maximum disorder).  
The two different shapes of brackets, $\langle\cdots\rangle$ and $[\cdots]$, denote 
the thermal average and the average over rewiring configurations, 
respectively.  A typical result is shown in Fig.~\ref{fig:binder}.  
For large $N$, there is a single crossing point $T_c$.  
It is noteworthy that the result is independent of 
the number of time slices $M$ once it exceeds a finite lower 
limit $M_c$, indicating that the correlation length $\xi_T$ 
in the time direction is a finite fraction of $\beta$.  When $T/J\gtrsim1$, 
the value of $M_c$ is typically $\sim30$ and it increases 
at lower temperatures.

\begin{figure}[bth]
\centering
\includegraphics[width=8cm]{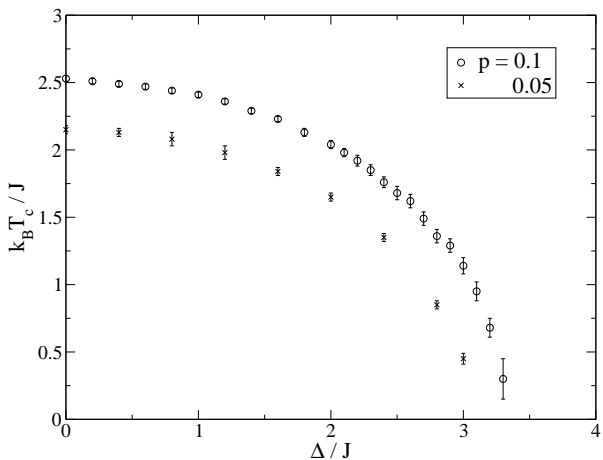}
\caption{Phase boundary of a transverse-field Ising model on a
  small-world network.  The small-world network was constructed
  following Watts and Strogatz starting from the one-dimensional regular
  network with $k=2$ and rewiring the links with the probability
  $p=0.1$ and 0.05.}
\label{fig:phase_diagram}
\end{figure}

Figure~\ref{fig:phase_diagram} shows phase diagrams in $T$-$\Delta$ space 
for $k=2$ and two different rewiring probabilities, $p=0.1$ and $0.05$.  
First of all, we find $T_c>0$ in the absence of the 
transverse field, which agrees 
with previous results \cite{barrat00a,gitterman00a,herrero02a}.  
When a small $\Delta$ is turned on, $T_c$ remains finite 
although it decreases with increasing $\Delta$.  At a fixed $\Delta$, 
we find that $T_c$ increases with $p$, 
just as in the no field case.  

Extrapolating the phase transition line to $T=0$, 
one may obtain the quantum critical transverse field $\Delta_c$.  
This is an extension of the quantum critical point 
of the regular Ising model \cite{sachdev98booka}.
From Fig.~\ref{fig:phase_diagram}, one can clearly 
see that $\Delta_c$ increases with $p$.  
This implies that the more the Ising system is rewired, the 
more resilient it is to quantum fluctuations.

\begin{figure}[bth]
\centering
\includegraphics[width=8cm]{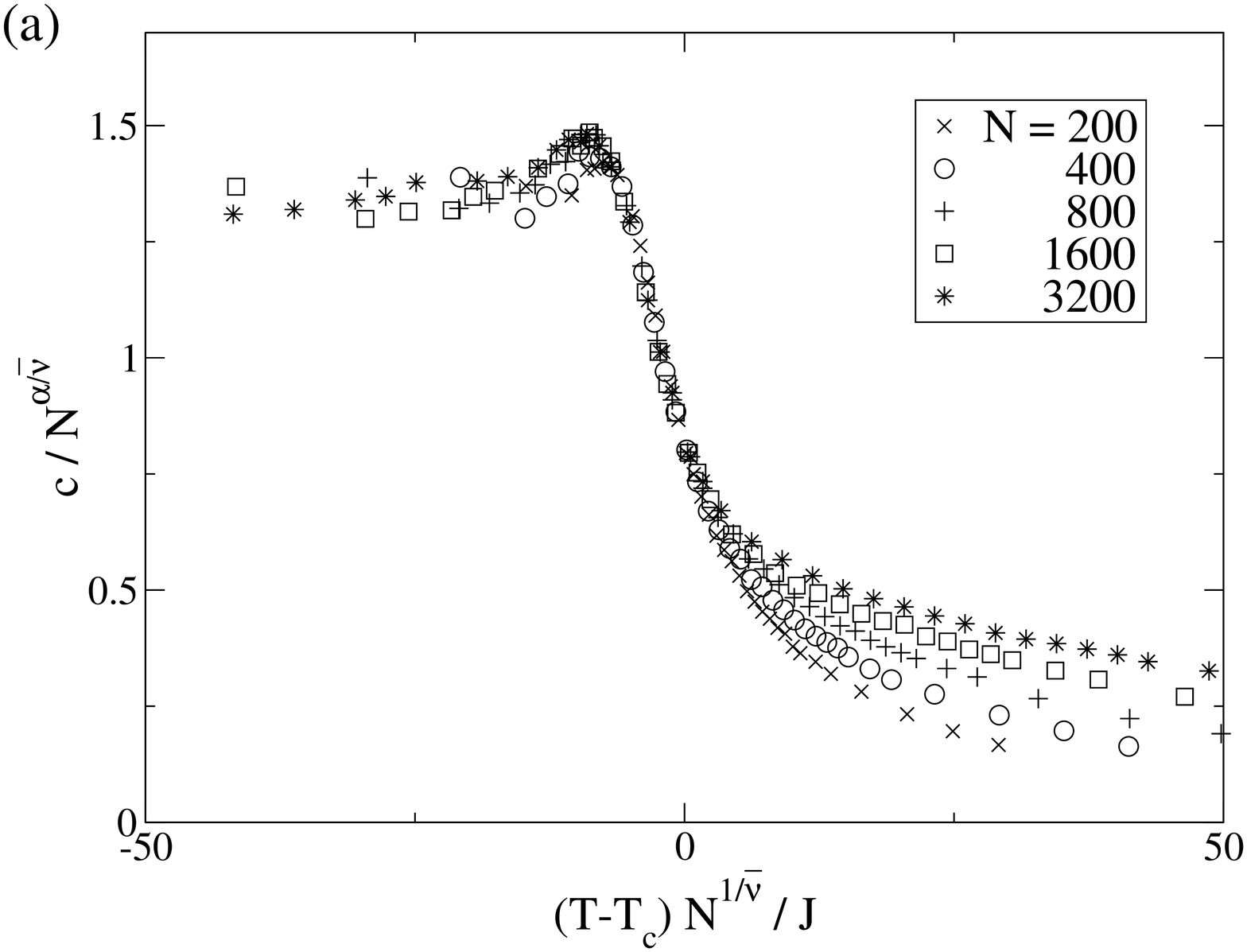}
\includegraphics[width=8cm]{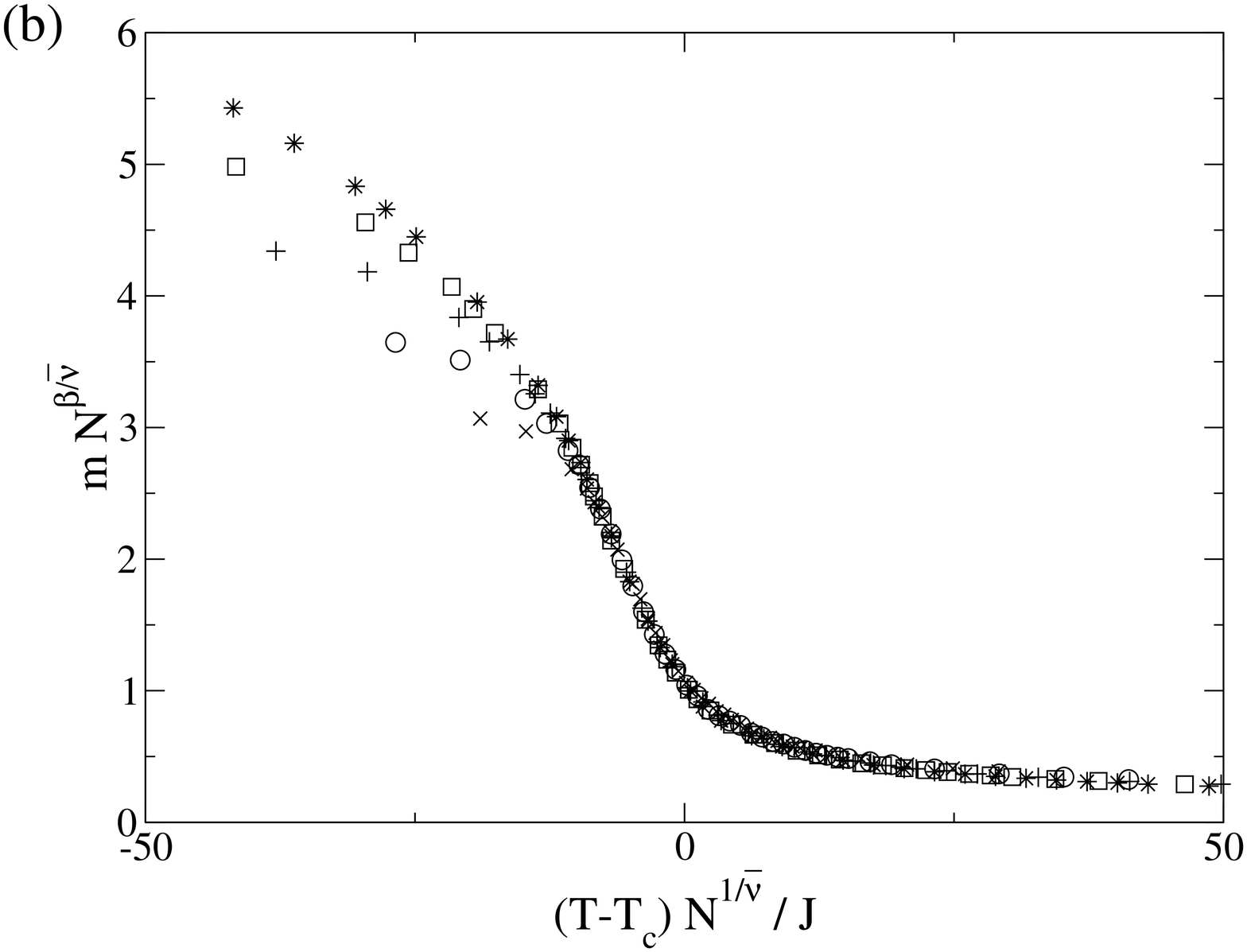}
\includegraphics[width=8cm]{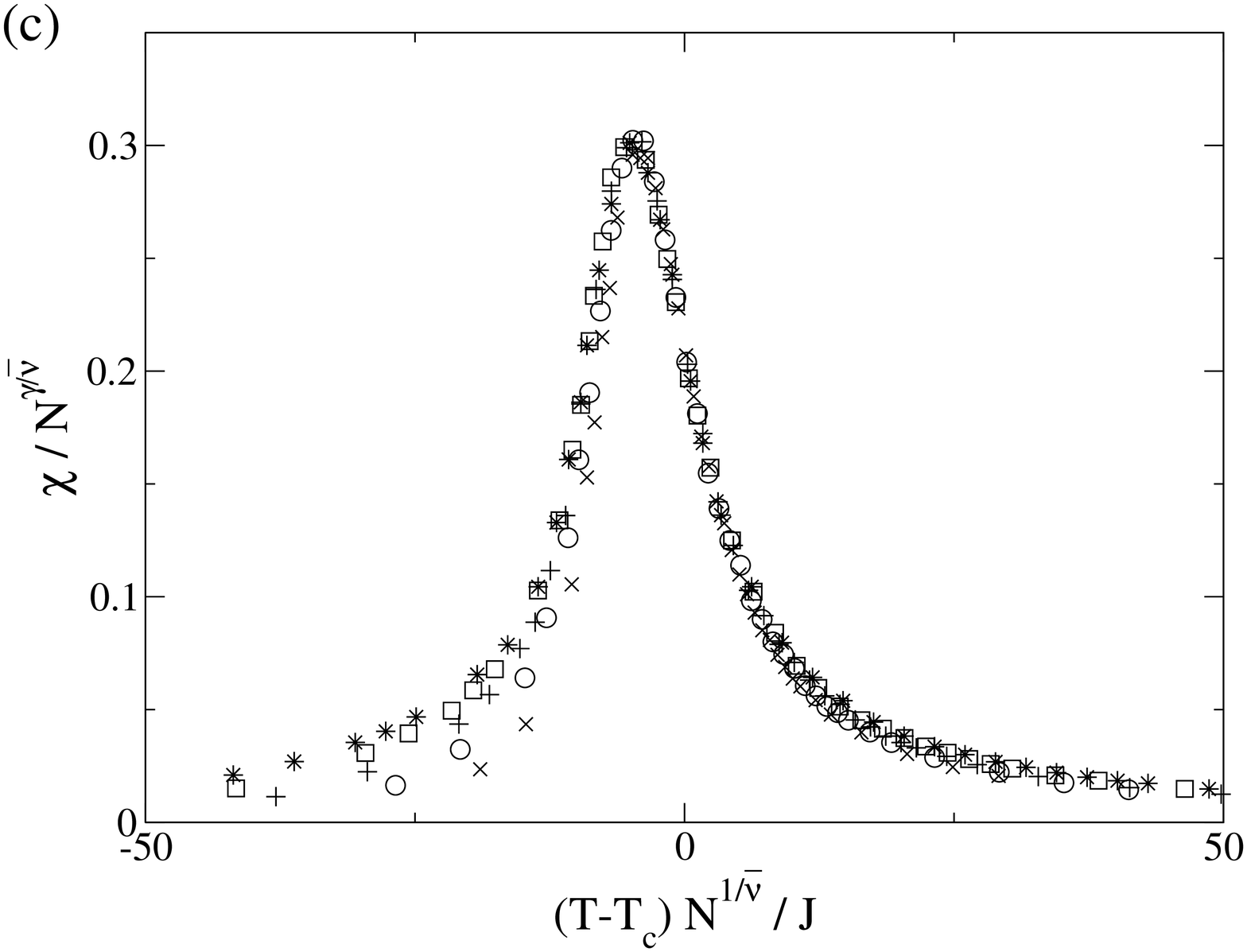}
\caption{Universal scaling functions 
at $k=2$, $p=0.1$, $B/J=1$, $T_c/J=2.04$, and $M=30$: 
(a) specific heat, 
(b) magnetization, and (c) susceptibility.  Each quantity 
is measured per site.  
The legend in (a) is common to all three figures.}
\label{fig:scaling}
\end{figure}

We now characterize the ferromagnetic phase transition 
by determining the universality class to which it belongs.  
In the pure classical case ($\Delta=0$), it is known that the 
phase transition is
mean-field like \cite{barrat00a,gitterman00a,kimBJ01a}.  To address the
question whether this is still the case for finite values of
the transverse field $\Delta$, we investigated the scaling behavior 
of the phase transition described above.  
First of all, the Binder cumulant was fitted to a 
scaling function of the form
\begin{equation}
U_N(T) = \tilde{U}\left((T-T_c)N^{1/\overline{\nu}}\right).
\end{equation}
Due to the infinite-range nature of the small-world 
network, the above exponent $\overline{\nu}$ describes 
the divergence of coherence number $N_c$ instead of 
the correlation length $\xi$ \cite{botet82a,kimBJ01a}.  More 
explicitly, we may write
\begin{equation}
N_c \propto |T-T_c|^{-\overline{\nu}},
\end{equation}
near the transition.

The other critical exponents have also been obtained from various physical
quantities by fitting them to scaling functions.  For example, the
specific heat per spin is fitted to
\begin{equation}
c(T) = N^{\alpha/\overline{\nu}}\tilde{c}
\left((T-T_c)N^{1/\overline{\nu}}\right) \,.
\end{equation}
Figure~\ref{fig:scaling} shows an example of the scaling functions for
(a) specific heat, (b) magnetization, and (c) susceptibility per spin.
If we choose appropriate exponents, results from systems of different
sizes clearly collapse to one single curve near $T_c$.  As 
$T$ moves away from the scaling regime, the curves deviate.  
The best fitting values of the exponents are summarized in 
Table~\ref{tab:1}.  It turned out that they are the same as those of
the mean-field transition to a very high precision.
Therefore, we conclude that the quantum fluctuations 
introduced by the transverse field do not alter 
the universality class of the ferromagnetic phase 
transition in the Ising models on small-world networks.

\begin{table}[bth]
\centering
\begin{math}
\begin{array}{llll}
\hline\hline
\text{quantity} & \text{critical behavior} & \text{our result} &
\text{MF value} \\
\hline
\text{specific heat} & c \propto |T-T_c|^{-\alpha} & \alpha =0.0
& \alpha = 0 \\
\text{magnetization} & m \propto (T_c-T)^\beta & \beta = 0.5
& \beta = 1/2 \\
\text{susceptibility} & \chi \propto |T-T_c|^{-\gamma} & \gamma = 1.0
& \gamma = 1
\\
\text{coherence number} & N_c \propto |T-T_c|^{-\bar\nu}
& \bar\nu = 2.0 & \bar\nu = 2 \\
\hline\hline
\end{array}
\end{math}
\caption{Critical exponents near the ferromagnetic phase transitions for
  $\Delta<\Delta_c$ in an transver-field Ising model on a small-world
  network.  For comparison, mean-field (MF) values are also provided.}
\label{tab:1}
\end{table}

So far, we have studied the scaling properties of an Ising system on
small-world networks at finite temperatures.  However, the quantum
critical point $\Delta_c$ at $T=0$ is in itself of much interest, since
in general the universality class of a quantum critical point is
different from that of classical transitions at finite
temperatures \cite{sachdev98booka}.  Whether the small-world network
will change the universality class of the quantum critical point as
compared to that in a regular Ising model is a highly intriguing
question.  In order to address that question, however, one has to use a
different technique than those used above, because the results from the
quantum Monte Carlo simulations become unreliable near $T=0$.  Therefore
we leave it as a topic for further study.

In summary, we have used quantum Monte Carlo simulations 
to obtain the phase diagram of 
an Ising system on small-world networks in the presence of 
a transverse magnetic field.  
The ferromagnetic phase persisted 
at finite $\Delta$, although the effect of quantum fluctuations 
introduced by the transverse field was manifested by the 
decrease of $T_c$.  At a fixed $\Delta$, we have also shown 
that $T_c$ increases with the rewiring probability 
$p$.  From various scaling exponents, 
we have argued that the ferromagnetic phase transition 
at finite field was still of a mean-field type.  
Eventually, $T_c$ decreases to zero at a quantum critical 
point at a finite field $\Delta_c$, but 
$\Delta_c$ increases with increasing $p$.  Since the 
ferromagnetic region increases with $p$ in both $T$ and 
$\Delta$ directions, we conclude that the small-world topology 
of the spin system competes against both thermal and quantum 
fluctuations and enhances the correlation and ordering of the spins.

\acknowledgments
We thank M.~Y. Choi, H. Hong, and B.~J. Kim for very useful discussions.
We acknowledge the support from the Swiss-Korean Outstanding Research
Efforts Award program (SKORE-A).

\bibliography{sw_ising}

\begin{thebibliography}{23}
\bibitem[{uni()}]{uniform}
\bibinfo{note}{A more realistic model on solids, $J_{ij}$ may also depend on
  distance between $i$ and $j$. Here we adopt the simplest model.}
\bibitem[{k>1()}]{k>1}
\bibinfo{note}{We restrict $k\geq 2$ since the WS model with $k=1$ has zero
  clustering coefficient and is exceptional in many ways. See for example,
  Ref.~\onlinecite{barrat00a}.}
\end{thebibliography}

\end{document}